\documentclass[runningheads]{llncs}

\usepackage[numbers,sort&compress]{natbib}
\usepackage[linesnumbered,ruled,vlined]{algorithm2e}
\usepackage{amsmath,amssymb}
\usepackage{booktabs}
\usepackage{microtype}
\usepackage{listings}
\usepackage{xcolor}
\usepackage{float}
\usepackage{multirow}
\usepackage{array}
\usepackage{pifont}
\usepackage[section]{placeins}
\usepackage{pgfplots}
\pgfplotsset{compat=1.17}
\usepackage{hyperref}
\hypersetup{colorlinks=true, linkcolor=black, citecolor=black, urlcolor=black}
\usepackage{tikz}
\usetikzlibrary{shapes.geometric, arrows.meta, positioning, calc, backgrounds, fit}
\newcommand{\botrule}{\bottomrule}

\begin{document}
%% ================================================================

\title{SynQL: A Controllable and Scalable Rule-Based Framework for SQL Workload Synthesis for Performance Benchmarking
}

\author{Kahan Mehta\inst{1}\thanks{Corresponding author.} \and
        Amit Mankodi\inst{1}}

\institute{School of Technology, Dhirubhai Ambani University,
           Gandhinagar, Gujarat, India\\
           \email{202411038@dau.ac.in},\
           \email{amit\_mankodi@dau.ac.in}}

\authorrunning{K.\ Mehta and A.\ Mankodi}
\titlerunning{SynQL: Controllable SQL Workload Synthesis}

\maketitle

%% ================================================================
\begin{abstract}
Database research and the development of learned query optimisers rely
heavily on realistic SQL workloads. Acquiring real-world queries is
increasingly difficult, however, due to strict privacy regulations, and
publicly released anonymised traces typically strip out executable query text
to preserve confidentiality. Existing synthesis tools fail to bridge this
training-data gap: traditional benchmarks offer too few fixed templates for
statistical generalisation, while Large Language Model (LLM) approaches
suffer from \emph{schema hallucination}---fabricating non-existent
columns---and \emph{topological collapse}---systematically defaulting to
simplistic join patterns that fail to stress-test query optimisers.

We propose \textbf{SynQL}, a deterministic workload synthesis framework that
generates structurally diverse, execution-ready SQL workloads. As a
foundational step toward bridging the training-data gap, SynQL targets the
core SQL fragment---multi-table joins with projections, aggregations, and
range predicates---which dominates analytical workloads. SynQL abandons
probabilistic text generation in favour of traversing the live database's
foreign-key graph to populate an Abstract Syntax Tree (AST), guaranteeing
schema and syntactic validity by construction. A configuration vector
$\Theta$ provides explicit, parametric control over join topology (Star,
Chain, Fork), analytical intensity, and predicate selectivity. Experiments on
TPC-H and IMDb show that SynQL produces near-maximally diverse workloads
(Topological Entropy $H = 1.53$~bits) and that tree-based cost models trained
on the synthetic corpus achieve $R^{2} \geq 0.79$ on held-out synthetic test
sets with sub-millisecond inference latency, establishing SynQL as an
effective foundation for generating training data when production logs are
inaccessible.

\keywords{Query Execution Time \and SQL Workload Synthesis \and Learned Cost
          Models \and Topological Diversity \and Synthetic Benchmarks}
\end{abstract}

%% ================================================================
\section{Introduction}\label{sec:intro}
%% ================================================================

Modern cloud databases and learned query optimisers depend on large,
representative SQL workloads to guide system tuning, evaluate performance, and
train cost-estimation models. Obtaining such workloads from production systems
is increasingly difficult: privacy regulations, security protocols, and data
governance constraints prevent the research community from accessing queries
that contain sensitive intellectual property or personally identifiable
information. This challenge mirrors the broader data-access barriers observed
across domains---from healthcare to
finance---where synthetic data generation has emerged as a principled
alternative to real data when privacy constraints are
binding~\cite{abdulrahman2025synthetic}.

Cloud vendors have responded by releasing anonymised traces---Snowset~\cite{dageville2016snowflake} and
Redset~\cite{jain2024redset} are notable examples---but these artefacts retain only high-level
execution statistics (CPU time, bytes scanned) and deliberately omit the
original SQL text and underlying data. The traces are therefore not executable
and cannot be used to train machine learning models directly.

Researchers who wish to train learned database components are left with two
inadequate alternatives. The first is \emph{fixed benchmarks}: TPC-H and
TPC-DS provide 22 and 99 hand-crafted query templates,
respectively~\cite{poess2000tpc,nambiar2006making}. Training on so narrow a template set leads
to catastrophic overfitting and fails to capture the complex, long-tailed
distributions observed in production. The Join Order Benchmark
(JOB)~\cite{leis2015good} improves realism by introducing real-world data
skew, yet its template count remains small, and its 2025
revisitation~\cite{leis2024still} confirms that optimizer gains over the last
decade remain marginal for structurally atypical queries. The second
alternative is \emph{generative LLMs}: because LLMs treat SQL generation as a
probabilistic token-prediction task rather than a constrained graph-routing
problem, they exhibit schema hallucination (fabricating non-existent columns
or tables) and topological collapse (defaulting to simplistic hub-and-spoke
star joins that dominate pre-training corpora). Spider~2.0 documents a drop
in GPT-4o's success rate from 86.6\% on simple schemas to just 10.1\% on
enterprise SQL~\cite{lei2024spider2}, and BIRD confirms execution accuracy
below 40\% on complex schemas~\cite{li2023bird}.

To overcome these limitations we propose \textbf{SynQL}
(\textbf{Syn}thetic \textbf{Q}uery \textbf{L}anguage workload generator), a
fine-grained synthesis framework that replaces probabilistic generation with
deterministic schema-graph traversal. By constructing queries node-by-node
from the target database's live foreign-key catalog and assembling them into
an Abstract Syntax Tree (AST), SynQL guarantees 100\% schema and syntactic
validity by construction, mechanically bypassing the hallucination risks of
autoregressive models. A configuration vector $\Theta$ provides explicit
parametric control over join topology, analytical intensity, and predicate
selectivity, enabling practitioners to synthesise massive, structurally diverse
training datasets without hand-authoring a single template.

The remainder of the paper is organised as follows.
Section~\ref{sec:related} surveys related work.
Section~\ref{sec:framework} presents the SynQL framework, including its
two-phase pipeline, algorithms, and configuration parameters.
Section~\ref{sec:experiments} details the experimental setup and case studies.
Section~\ref{sec:results} reports results.
Section~\ref{sec:limitations} discusses limitations, and
Section~\ref{sec:conclusion} concludes.

%% ================================================================
\section{Related Work}\label{sec:related}
%% ================================================================

Our framework addresses limitations at the intersection of three active
research areas: benchmark coverage for learned systems, generalisation in
learned query optimisation, and the structural constraints of automated SQL
synthesis.

\subsection{The Training-Data Barrier: Industrial Needs
            vs.\ Static Benchmarks}\label{subsec:rw-benchmarks}

The industrial viability of tree-based query execution time (QET) prediction
is definitively demonstrated by Amazon's Stage predictor~\cite{wu2024stage}.
Deployed as a hierarchical XGBoost ensemble across Redshift instances, Stage
achieves a 20\% average latency reduction in production while making the
data-access barrier explicit: its local models are trained on massive logs of
customer-specific production queries that academic practitioners cannot
replicate. Standard benchmarks such as TPC-H and TPC-DS are structurally
insufficient substitutes; their fixed template sets are too narrow for
statistical generalisation. The JOB-Complex challenge~\cite{wehrstein2025job}
provides further evidence that structurally atypical queries---precisely those
absent from standard benchmarks---cause the most severe performance regressions
in modern systems.

\subsection{Learned Optimisation and Generalisation
            Bottlenecks}\label{subsec:rw-learned}

End-to-end QET prediction depends critically on cardinality estimation. The
field has evolved from early deep-learning estimators such as
MSCN~\cite{kipf2019} and NeuroCard~\cite{yang2020neurocard} through joint
plan-cost networks~\cite{sun2019end} and reinforcement-learning-based
optimisers such as Bao~\cite{bao2021} and Balsa~\cite{yang2022balsa}. Despite
these architectural advances, a universal limitation persists: model
performance generalises poorly to structurally novel queries absent from
training data. LIMAO~\cite{limao2025} attempts to address this through a
lifelong modular architecture, yet the survey by Zhu et
al.~\cite{zhu2024learned} concludes that training-data quality remains the
single most consequential factor in model generalisation. Sun et al.'s
comparative study~\cite{sun2021learnedcard} further shows that no single
estimator dominates across all execution regimes. SynQL directly targets this
bottleneck: the $P_{\text{where}}$ and $\alpha_{\text{shape}}$ parameters are
designed to generate the estimation regimes and topological edge cases that
trained models most often lack.

\subsection{Failure Modes of LLM-Based SQL
            Synthesis}\label{subsec:rw-llm}

The community has explored LLMs for workload synthesis, but empirical
benchmarks reveal severe mechanical limitations when tasked with structural SQL
generation. Spider~2.0~\cite{lei2024spider2} attributes GPT-4o's sharp
accuracy drop on enterprise SQL to schema hallucination and dialect confusion.
Scale-driven evaluations in BIRD~\cite{li2023bird} and
PARROT~\cite{zhou2025parrotbenchmarkevaluatingllms} confirm execution accuracy
below 40\% on complex schemas. The survey by Hong et
al.~\cite{hong2025nextgen} identifies two persistent failure modes in
autoregressive SQL generation: schema compliance failures and structural
rigidity driven by training-corpus bias. Because hallucination and topological
collapse rates are unacceptably high for automated data generation, SynQL is
proposed as a constructive, graph-based alternative that mathematically
bypasses these flaws to guarantee strict validity.

%% ================================================================
\section{The SynQL Framework}\label{sec:framework}

This section presents the complete SynQL design.
Section~\ref{subsec:overview} provides an architectural overview and
introduces the notation used throughout.
Sections~\ref{subsec:phase1}~and~\ref{subsec:phase2} detail the two
core algorithms.
Section~\ref{subsec:generation_loop} describes how these algorithms
compose into the workload generation loop.
Section~\ref{subsec:theta} explains the configuration
vector~$\Theta$, and Section~\ref{subsec:case_studies} illustrates the
framework with a concrete example.

% ------------------------------------------------------------------
\subsection{Architectural Overview}\label{subsec:overview}
% ------------------------------------------------------------------

\begin{figure}[tb]
\centering
\resizebox{\columnwidth}{!}{%
\begin{tikzpicture}[
    >=Stealth,
    node distance=0.8cm and 0.7cm,
    inputbox/.style={
        draw, rounded corners=4pt, fill=blue!8,
        minimum width=2.0cm, minimum height=1.4cm,
        font=\small, align=center, text width=1.9cm,
        line width=0.6pt
    },
    phasebox/.style={
        draw, rounded corners=4pt, fill=orange!12,
        minimum width=2.4cm, minimum height=1.4cm,
        font=\small\bfseries, align=center, text width=2.3cm,
        line width=0.8pt
    },
    midbox/.style={
        draw, rounded corners=4pt, fill=violet!8,
        minimum width=2.0cm, minimum height=1.4cm,
        font=\small, align=center, text width=1.9cm,
        line width=0.6pt
    },
    outputbox/.style={
        draw, rounded corners=4pt, fill=green!10,
        minimum width=2.0cm, minimum height=1.4cm,
        font=\small, align=center, text width=1.9cm,
        line width=0.6pt
    },
    configbox/.style={
        draw, rounded corners=3pt, fill=gray!10,
        minimum width=1.8cm, minimum height=0.65cm,
        font=\scriptsize, align=center,
        line width=0.5pt
    },
    detailbox/.style={
        draw=none, fill=none,
        font=\scriptsize\itshape, align=center, text width=2.2cm
    },
    arrow/.style={->, thick, >=Stealth, line width=0.7pt},
    dataarrow/.style={->, thick, >=Stealth, line width=0.7pt,
                      color=blue!60},
]

% === Main pipeline (left to right) ===
\node[inputbox] (catalog)
  {Database\\Catalog\\[-2pt]
   {\scriptsize $\mathcal{S}\!=\!(\mathcal{V},\mathcal{E})$}};

\node[phasebox, right=1.0cm of catalog] (phase1)
  {Phase~I\\[-2pt]Topological\\[-2pt]Traversal\\[-2pt]
   {\scriptsize (Alg.~\ref{alg:phase1})}};

\node[midbox, right=1.0cm of phase1] (blueprint)
  {Join\\Blueprint\\[-2pt]
   {\scriptsize $(\mathcal{T}_{\text{used}},\, J_q)$}};

\node[phasebox, right=1.0cm of blueprint] (phase2)
  {Phase~II\\[-2pt]Semantic Inj.\\[-2pt]+ AST Assembly\\[-2pt]
   {\scriptsize (Alg.~\ref{alg:phase2})}};

\node[outputbox, right=1.0cm of phase2] (query)
  {SQL Query\\$q_i$};

\node[outputbox, right=0.9cm of query, fill=green!18,
      line width=0.9pt] (workload)
  {Workload\\$\mathcal{Q}$\\[-2pt]
   {\scriptsize ($N$ queries)}};

% === Config Theta (above, spanning both phases) ===
\node[configbox, above=0.7cm of blueprint] (theta)
  {Config $\Theta$:\;
   $\alpha_{\text{shape}},\;K_{\text{join}},\;
    P_{\text{agg}},\;P_{\text{where}},\;K_{\text{pred}}$};

% === Detail annotations (below each phase) ===
\node[detailbox, below=0.15cm of phase1]  (d1)
  {FK-graph walk,\\$\alpha_{\text{shape}}$-weighted\\edge selection};
\node[detailbox, below=0.15cm of phase2]  (d2)
  {Column sampling,\\aggregation, predicates,\\AST compilation};

% === Arrows ===
\draw[arrow] (catalog)   -- (phase1);
\draw[dataarrow] (phase1)    -- (blueprint);
\draw[arrow] (blueprint) -- (phase2);
\draw[dataarrow] (phase2)    -- (query);
\draw[arrow] (query)     -- (workload);

\draw[arrow, gray, dashed]
  (theta.south) -- ++(0,-0.25) -| ([yshift=0.1cm]phase1.north);
\draw[arrow, gray, dashed]
  (theta.south) -- ++(0,-0.25) -| ([yshift=0.1cm]phase2.north);

% === Loop-back arrow (below) ===
\draw[arrow, dashed, gray!70, line width=0.6pt]
  ([yshift=-0.9cm]query.south) -- ++(0, -0.25)
  -| ([yshift=-1.15cm]phase1.south)
  -- ([yshift=-0.9cm]phase1.south)
  node[pos=0.45, below, font=\scriptsize\color{gray!70}]
       {repeat $N$ times};

\end{tikzpicture}%
}% end resizebox
\caption{SynQL pipeline overview.  The database catalog feeds Phase~I
  (Algorithm~\ref{alg:phase1}), which produces a join blueprint under
  topology bias~$\alpha_{\text{shape}}$.  Phase~II
  (Algorithm~\ref{alg:phase2}) injects semantic content and compiles
  each query via an AST\@.  Configuration vector~$\Theta$ governs both
  phases; the outer loop repeats them $N$~times to emit
  workload~$\mathcal{Q}$.}
\label{fig:arch}
\end{figure}
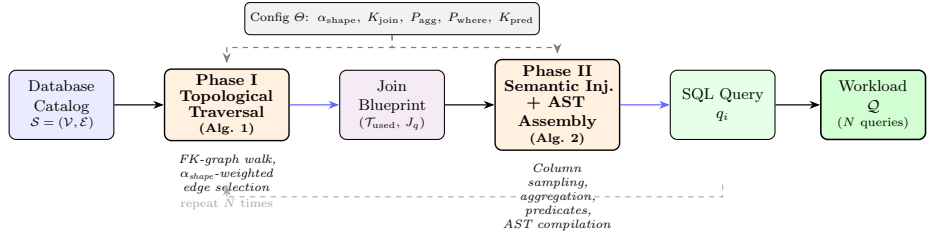

SynQL is a deterministic, two-phase constructive pipeline framework that bypasses
text-based probabilistic generation entirely.  Given a target database's
live catalog, the framework operates as follows
(Figure~\ref{fig:arch}):

\begin{enumerate}
  \item \textbf{Phase~I (Topological Traversal).}  The relational schema
        graph is traversed to produce a join subgraph whose shape is
        controlled by the topology-bias parameter~$\alpha_{\text{shape}}$.
        The output is a \emph{join blueprint}: the set of active tables
        and the edges connecting them.
  \item \textbf{Phase~II (Semantic Injection \& AST Assembly).}
        Projections, aggregations, and predicates are injected into the
        blueprint and compiled into an Abstract Syntax Tree (AST),
        producing an executable SQL string with guaranteed schema and
        syntactic validity.
\end{enumerate}

\noindent
To generate a workload of~$N$ queries, SynQL simply repeats Phase~I
and Phase~II in sequence~$N$~times under a shared configuration
vector~$\Theta$. Table~\ref{tab:notation} summarises all symbols and operations used in SynQL.

\begin{table}[p]
\centering
\caption{Notation and terminology used in the SynQL
  algorithms.}\label{tab:notation}
\small
\begin{tabular}{@{} l p{8.8cm} @{}}
\toprule
\textbf{Symbol / Term} & \textbf{Description} \\
\midrule
\multicolumn{2}{@{}l}{\textit{Schema and graph terms}} \\
$\mathcal{S} = (\mathcal{V}, \mathcal{E})$
  & Schema graph: $\mathcal{V}$ is the set of tables, $\mathcal{E}$ is
    the set of primary-key to foreign-key (PK-FK) edges \\
$T_{\text{base}}$
  & Root table selected uniformly at random to start the join traversal \\
$\mathcal{T}_{\text{used}}$
  & Set of tables visited so far during graph traversal \\
$J_q$
  & Set of join edges selected for the current query \\
$J_{\text{poss}}$
  & Candidate edge set: all non-cyclic PK-FK edges from visited to
    unvisited tables \\
$T_{\text{in}},\; T_{\text{out}}$
  & For a candidate edge~$e$: $T_{\text{in}}$ is the already-visited
    (anchor) table; $T_{\text{out}}$ is the unvisited table \\
$dist(v, T_{\text{base}})$
  & Shortest-path edge distance from table~$v$ to the root \\
$D_{\max}$
  & Current maximum depth:
    $\max_{v \in \mathcal{T}_{\text{used}}} dist(v, T_{\text{base}})$ \\
$w(e)$
  & Selection weight assigned to candidate edge~$e$
    (Eq.~\ref{eq:weight}) \\
$\mathbb{I}(\cdot)$
  & Indicator function: returns 1 if the condition is true, 0
    otherwise \\
\midrule
\multicolumn{2}{@{}l}{\textit{Configuration parameters
  ($\Theta$)}} \\
$\alpha_{\text{shape}}$
  & Topology bias: steers traversal toward Star
    ($\alpha_{\text{shape}}\!\to\!1$) or Chain
    ($\alpha_{\text{shape}}\!\to\!0$) topologies \\
$K_{\text{join}}$
  & Maximum number of join edges per query (controls table count) \\
$N_{\text{join}}$
  & Sampled join depth for one query:
    $N_{\text{join}} \sim U(1, K_{\text{join}})$ \\
$P_{\text{agg}}$
  & Probability that a numeric column is wrapped in an aggregation
    function (\texttt{SUM} or \texttt{AVG}) \\
$P_{\text{where}}$
  & Probability that a \texttt{WHERE} clause is added to the query \\
$K_{\text{pred}}$
  & Maximum number of predicates in the \texttt{WHERE} clause \\
\midrule
\multicolumn{2}{@{}l}{\textit{Query-construction terms}} \\
$C_{\text{select}}$
  & Accumulated \texttt{SELECT}-list entries (columns or aggregated
    expressions) \\
$C_{\text{group\_by}}$
  & Set of non-aggregated columns requiring a \texttt{GROUP BY} entry \\
$P_{\text{filters}}$
  & Set of generated predicate expressions for the \texttt{WHERE}
    clause \\
$\text{has\_agg}$
  & Boolean flag: \texttt{True} if at least one aggregation has been
    injected \\
$AST$
  & Abstract Syntax Tree representing the SQL query under
    construction \\
\midrule
\multicolumn{2}{@{}l}{\textit{Operations}} \\
$\mathrm{UniformRandom}(A)$
  & Draw one element uniformly at random from set~$A$ \\
$\mathrm{WeightedRandomChoice}(S, w)$
  & Draw one element from set~$S$ with probability proportional to
    weights~$w$ \\
$\mathrm{SampleColumns}(T)$
  & Return a random subset of catalog columns from table~$T$ \\
$\mathrm{SampleCatalogDomain}(c)$
  & Sample a domain-valid value for column~$c$ from database
    statistics \\
$\mathrm{IsNumeric}(c)$
  & Return \texttt{True} if column~$c$ has a numeric data type \\
$\mathrm{AggFunc}(c)$
  & Wrap column~$c$ in a randomly chosen aggregation function
    (\texttt{SUM} or \texttt{AVG}) \\
$\mathrm{CompileToSQLString}(AST)$
  & Serialize the AST into an executable SQL string \\
\bottomrule
\end{tabular}
\end{table}

% ---------- TikZ Architecture Diagram (Horizontal) ----------

% ------------------------------------------------------------------
\subsection{Phase~I: Topological Traversal}\label{subsec:phase1}
% ------------------------------------------------------------------

The goal of Phase~I is to construct a valid relational subgraph whose
shape reflects the topology bias dictated by
$\alpha_{\text{shape}}$.  Algorithm~\ref{alg:phase1} formalises the
procedure.

\paragraph{Schema Graph and Initialisation.}
SynQL ingests the database catalog to instantiate the relational schema
graph $\mathcal{S} = (\mathcal{V}, \mathcal{E})$, where $\mathcal{V}$
represents tables and $\mathcal{E}$ represents PK-FK constraints.  A
root table $T_{\text{base}}$ is sampled uniformly at random and a target
join depth $N_{\text{join}} \sim U(1, K_{\text{join}})$ is drawn.

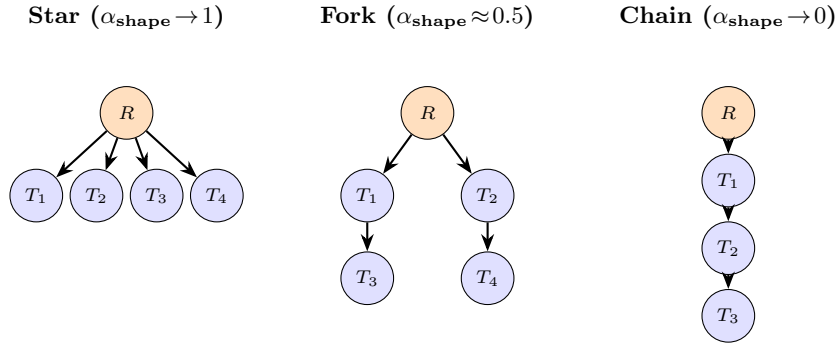
\begin{figure}[tb]
\centering
\begin{tikzpicture}[
    >=Stealth,
    tbl/.style={draw, circle, minimum size=0.7cm, inner sep=0pt,
                font=\scriptsize\bfseries},
    root/.style={tbl, fill=orange!25},
    leaf/.style={tbl, fill=blue!12},
    arrow/.style={->, thick},
    label/.style={font=\small\bfseries, above},
]

% --- Star (alpha -> 1) ---
\begin{scope}[shift={(0,0)}]
  \node[label] at (0, 1)
    {Star ($\alpha_{\text{shape}} \!\to\! 1$)};
  \node[root] (s0) at (0,0) {$R$};
  \node[leaf] (s1) at (-1.2, -1.1) {$T_1$};
  \node[leaf] (s2) at (-0.4, -1.1) {$T_2$};
  \node[leaf] (s3) at ( 0.4, -1.1) {$T_3$};
  \node[leaf] (s4) at ( 1.2, -1.1) {$T_4$};
  \draw[arrow] (s0) -- (s1);
  \draw[arrow] (s0) -- (s2);
  \draw[arrow] (s0) -- (s3);
  \draw[arrow] (s0) -- (s4);
\end{scope}

% --- Fork (alpha ~ 0.5) ---
\begin{scope}[shift={(4,0)}]
  \node[label] at (0, 1)
    {Fork ($\alpha_{\text{shape}} \!\approx\! 0.5$)};
  \node[root] (f0) at (0,0) {$R$};
  \node[leaf] (f1) at (-0.8, -1.1) {$T_1$};
  \node[leaf] (f2) at ( 0.8, -1.1) {$T_2$};
  \node[leaf] (f3) at (-0.8, -2.2) {$T_3$};
  \node[leaf] (f4) at ( 0.8, -2.2) {$T_4$};
  \draw[arrow] (f0) -- (f1);
  \draw[arrow] (f0) -- (f2);
  \draw[arrow] (f1) -- (f3);
  \draw[arrow] (f2) -- (f4);
\end{scope}

% --- Chain (alpha -> 0) ---
\begin{scope}[shift={(8,0)}]
  \node[label] at (0, 1)
    {Chain ($\alpha_{\text{shape}} \!\to\! 0$)};
  \node[root] (c0) at (0,  0)    {$R$};
  \node[leaf] (c1) at (0, -0.9)  {$T_1$};
  \node[leaf] (c2) at (0, -1.8)  {$T_2$};
  \node[leaf] (c3) at (0, -2.7)  {$T_3$};
  \draw[arrow] (c0) -- (c1);
  \draw[arrow] (c1) -- (c2);
  \draw[arrow] (c2) -- (c3);
\end{scope}

\end{tikzpicture}
\caption{Effect of $\alpha_{\text{shape}}$ on join topology.
  High values attach all tables to the root~$R$ (Star); low values
  extend the deepest frontier (Chain); intermediate values produce
  branching Forks.}
\label{fig:topologies}
\end{figure}

\paragraph{Iterative Edge Selection.}
At each expansion step the algorithm identifies~$J_{\text{poss}}$, the
set of all valid, non-cyclic PK-FK edges extending from
already-visited tables to unvisited tables.  For a candidate edge~$e$
connecting an active anchor $T_{\text{in}} \in \mathcal{T}_{\text{used}}$
to an unvisited table
$T_{\text{out}} \notin \mathcal{T}_{\text{used}}$, the selection weight
is:
\begin{equation}
  w(e) =
  \begin{cases}
    1 & \text{if } D_{\max} = 0, \\[6pt]
    \alpha_{\text{shape}} \cdot \mathbb{I}(T_{\text{in}} = T_{\text{base}})
    + (1 - \alpha_{\text{shape}}) \cdot
      \dfrac{dist(T_{\text{in}},\, T_{\text{base}})}{D_{\max}}
    & \text{if } D_{\max} > 0,
  \end{cases}
  \label{eq:weight}
\end{equation}
\noindent
During the first join step ($D_{\max}=0$) all candidate edges receive
equal weight to ensure unbiased root expansion.  On subsequent steps,
the bias parameter steers the topology as illustrated in
Figure~\ref{fig:topologies}:
\begin{itemize}
  \item $\alpha_{\text{shape}} \to 1$: heavily weights edges returning to
        the root, producing wide \textbf{Star} topologies;
  \item $\alpha_{\text{shape}} \to 0$: favours deepest-frontier expansion,
        producing deep \textbf{Chain} topologies;
  \item Intermediate values yield \textbf{Fork} (hybrid) topologies.
\end{itemize}
Because the traversal follows actual FK constraints, every generated join
is semantically valid by construction.

% ---------- TikZ Topology Illustration ----------

%%% --- Algorithm 1: Phase I ---
\begin{algorithm}[t]
\caption{Phase~I: Topological Graph Traversal.  All symbols are
  defined in Table~\ref{tab:notation}.}\label{alg:phase1}
\SetAlgoLined
\DontPrintSemicolon
\KwIn{Schema graph $\mathcal{S}=(\mathcal{V}, \mathcal{E})$,
      max join depth $K_{\text{join}}$,
      topology bias $\alpha_{\text{shape}}$}
\KwOut{Active table set $\mathcal{T}_{\text{used}}$,
      join edge set $J_q$}

$T_{\text{base}} \leftarrow
  \mathrm{UniformRandom}(\mathcal{V})$\;
$\mathcal{T}_{\text{used}} \leftarrow \{T_{\text{base}}\}$;\quad
$J_q \leftarrow \emptyset$\;
$N_{\text{join}} \leftarrow
  \mathrm{UniformRandom}(1, K_{\text{join}})$
  \tcp*{target join count}
\While{$|J_q| < N_{\text{join}}$}{
  $J_{\text{poss}} \leftarrow \{ (T_a, T_b) \in \mathcal{E} \mid
    (T_a \in \mathcal{T}_{\text{used}}) \oplus
    (T_b \in \mathcal{T}_{\text{used}}) \}$
    \tcp*{non-cyclic FK edges}
  \lIf{$J_{\text{poss}} = \emptyset$}{\textbf{break}
    \tcp*{schema exhausted}}
  $D_{\max} \leftarrow
    \max_{v \in \mathcal{T}_{\text{used}}}
    dist(v, T_{\text{base}})$
    \tcp*{current max depth}
  \For{each $e \in J_{\text{poss}}$}{
    $T_{\text{in}} \leftarrow e \cap
      \mathcal{T}_{\text{used}}$
      \tcp*{anchor table}
    Compute $w(e)$ via Eq.~(\ref{eq:weight}) using
      $T_{\text{in}},\, T_{\text{base}},\, D_{\max},\,
       \alpha_{\text{shape}}$\;
  }
  $e^{*} = (T_a, T_b) \leftarrow
    \mathrm{WeightedRandomChoice}(J_{\text{poss}}, w)$\;
  $J_q \leftarrow J_q \cup \{e^{*}\}$;\quad
  $\mathcal{T}_{\text{used}} \leftarrow
    \mathcal{T}_{\text{used}} \cup \{T_a, T_b\}$\;
}
\Return $\mathcal{T}_{\text{used}},\; J_q$
\end{algorithm}

% ------------------------------------------------------------------
\subsection{Phase~II: Semantic Injection and AST
            Assembly}\label{subsec:phase2}
% ------------------------------------------------------------------

Given the join blueprint $(\mathcal{T}_{\text{used}}, J_q)$ produced by
Phase~I, Phase~II populates the query with semantic content and compiles
it into an executable SQL string.  Algorithm~\ref{alg:phase2} formalises
the three sub-steps.

\paragraph{Step~1: Analytical Injection.}
For each table $T \in \mathcal{T}_{\text{used}}$, SynQL samples a random
subset of catalog columns.  Each numeric column is wrapped in an
aggregation function (\texttt{SUM} or \texttt{AVG}) with probability
$P_{\text{agg}}$.  Any column that is \emph{not} aggregated is
automatically enrolled in the $C_{\text{group\_by}}$ set.  This
invariant guarantees syntactic validity: every non-aggregated column in
the \texttt{SELECT} list will have a corresponding \texttt{GROUP BY}
entry, mechanically eliminating the ``unaggregated column'' errors that
plague LLM-based generators.

\paragraph{Step~2: Predicate Injection.}
With probability $P_{\text{where}}$, a \texttt{WHERE} clause is added.
Up to $K_{\text{pred}}$ range or equality predicates are generated using
domain-aware values sampled from the database's statistics catalog
(e.g.,~$c > \mathit{val}$), ensuring that predicates reference valid
column domains.

\paragraph{Step~3: AST Compilation.}
The accumulated state---projections, join edges, predicates, and
grouping columns---is mapped into a relational AST that enforces strict
SQL grammar rules, making syntactic errors mechanically impossible.  If
$C_{\text{group\_by}}$ is non-empty and at least one aggregation is
present, a \texttt{GROUP BY} clause is automatically appended.  Optional
\texttt{ORDER BY} and \texttt{LIMIT} clauses are attached before the
AST is compiled into an SQL string.

%%% --- Algorithm 2: Phase II ---
\begin{algorithm}[t]
\caption{Phase~II: Semantic Injection and AST Assembly.  All symbols
  are defined in Table~\ref{tab:notation}.}\label{alg:phase2}
\SetAlgoLined
\DontPrintSemicolon
\KwIn{Join blueprint $(\mathcal{T}_{\text{used}}, J_q)$ from
      Algorithm~\ref{alg:phase1},\;
      configuration parameters
      $P_{\text{agg}}, P_{\text{where}}, K_{\text{pred}}$}
\KwOut{Executable SQL query $q$}

$C_{\text{select}} \leftarrow \emptyset$;\quad
$C_{\text{group\_by}} \leftarrow \emptyset$;\quad
$P_{\text{filters}} \leftarrow \emptyset$;\quad
$\text{has\_agg} \leftarrow \text{False}$\;
\BlankLine
\tcp{Step 1: Analytical Injection}
\For{each $T \in \mathcal{T}_{\text{used}}$}{
  \For{each $c \in \mathrm{SampleColumns}(T)$}{
    \eIf{$\mathrm{IsNumeric}(c) \land \mathrm{Rand}() < P_{\text{agg}}$}{
      $C_{\text{select}} \leftarrow
        C_{\text{select}} \cup \{\mathrm{AggFunc}(c)\}$\;
      $\text{has\_agg} \leftarrow \text{True}$\;
    }{
      $C_{\text{select}} \leftarrow
        C_{\text{select}} \cup \{c\}$\;
      $C_{\text{group\_by}} \leftarrow
        C_{\text{group\_by}} \cup \{c\}$
        \tcp*{enforce GROUP BY invariant}
    }
  }
}
\BlankLine
\tcp{Step 2: Predicate Injection}
\If{$\mathrm{Rand}() < P_{\text{where}}$}{
  \For{each $c \in \mathrm{SampleColumns}(\mathcal{T}_{\text{used}},\,
                                          K_{\text{pred}})$}{
    $\mathit{val} \leftarrow \mathrm{SampleCatalogDomain}(c)$\;
    $P_{\text{filters}} \leftarrow
      P_{\text{filters}} \cup \{c > \mathit{val}\}$\;
  }
}
\BlankLine
\tcp{Step 3: AST Compilation}
$AST \leftarrow \mathrm{InitNode}(\texttt{SELECT},\;
  C_{\text{select}})$\;
$\mathrm{AttachNode}(AST,\; \texttt{FROM},\;
  \mathcal{T}_{\text{used}},\; J_q)$\;
\lIf{$P_{\text{filters}} \neq \emptyset$}{%
  $\mathrm{AttachNode}(AST,\; \texttt{WHERE},\;
    P_{\text{filters}})$}
\lIf{$\text{has\_agg} \land
  C_{\text{group\_by}} \neq \emptyset$}{%
  $\mathrm{AttachNode}(AST,\; \texttt{GROUP\_BY},\;
    C_{\text{group\_by}})$}
Optionally attach \texttt{ORDER BY} and \texttt{LIMIT} clauses\;
$q \leftarrow \mathrm{CompileToSQLString}(AST)$\;
\Return $q$
\end{algorithm}

% ------------------------------------------------------------------
\subsection{Workload Generation Loop}\label{subsec:generation_loop}
% ------------------------------------------------------------------

Algorithms~\ref{alg:phase1}~and~\ref{alg:phase2} together produce a
single executable query.  To generate a full workload~$\mathcal{Q}$ of
$N$~queries, SynQL executes them in sequence inside a simple outer
loop:

\begin{enumerate}
  \item \textbf{Reset.}  All per-query state
        ($\mathcal{T}_{\text{used}}, J_q, C_{\text{select}},
        C_{\text{group\_by}}, P_{\text{filters}}, \text{has\_agg}$) is
        cleared.
  \item \textbf{Phase~I.}  Algorithm~\ref{alg:phase1} is invoked with
        the shared configuration parameters
        $(\alpha_{\text{shape}}, K_{\text{join}})$ to produce a fresh
        join blueprint.
  \item \textbf{Phase~II.}  Algorithm~\ref{alg:phase2} receives the
        blueprint and the remaining parameters
        $(P_{\text{agg}}, P_{\text{where}}, K_{\text{pred}})$ to emit
        query~$q_i$.
  \item \textbf{Enqueue.}  $q_i$ is appended to~$\mathcal{Q}$.
\end{enumerate}

\noindent
Because each iteration draws a new root table uniformly at random,
schema-wide table coverage is ensured over large workloads.  The
procedure runs in $O(N \cdot K_{\text{join}})$ time; the full
20{,}000-query corpus used in our experiments
(Section~\ref{sec:experiments}) was generated in under ten minutes on
the target hardware.

\begin{figure}[p]
\centering
\begin{minipage}[c]{0.60\textwidth}
\centering
\resizebox{\textwidth}{!}{%
\begin{tikzpicture}[
    >=Stealth,
    node distance=0.45cm,
    stepbox/.style={
        draw, rounded corners=3pt, fill=#1,
        minimum height=0.75cm, minimum width=5.8cm,
        font=\small, align=left, text width=5.6cm,
        inner sep=5pt, line width=0.5pt
    },
    stepbox/.default={white},
    phaselabel/.style={
        draw, rounded corners=4pt, fill=#1,
        minimum height=0.55cm, minimum width=5.8cm,
        font=\small\bfseries, align=center,
        line width=0.7pt
    },
    phaselabel/.default={orange!15},
    grapharea/.style={
        draw=gray!70, dashed, rounded corners=3pt, fill=orange!4,
        minimum height=2.0cm, minimum width=5.8cm,
        inner sep=6pt
    },
    tblnode/.style={
        draw, circle, minimum size=0.65cm, inner sep=0pt,
        font=\scriptsize\bfseries, fill=#1,
        line width=0.6pt
    },
    tblnode/.default={blue!25},
    rootnode/.style={tblnode=orange!40},
    sqlbox/.style={
        draw, rounded corners=3pt, fill=green!6,
        font=\ttfamily\scriptsize, align=left, text width=5.4cm,
        inner sep=5pt, line width=0.6pt,
        minimum width=5.8cm
    },
    outputlabel/.style={
        draw, rounded corners=4pt, fill=green!15,
        minimum height=0.55cm, minimum width=5.8cm,
        font=\small\bfseries, align=center,
        line width=0.7pt
    },
    arrow/.style={->, thick, >=Stealth, line width=0.6pt},
    steplabel/.style={
        font=\scriptsize\bfseries, text=gray!60,
        anchor=east
    },
]

% ======== PHASE I HEADER ========
\node[phaselabel=orange!15] (p1hdr)
  {Phase~I: Topological Traversal (Alg.~\ref{alg:phase1})};

% Step 1
\node[stepbox=blue!6, below=0.35cm of p1hdr] (s1)
  {\textbf{1.\ Select root table}\quad
   $T_{\text{base}} \leftarrow \mathrm{UniformRandom}(\mathcal{V})$
   \quad$\Rightarrow$\quad \texttt{title}};

% Step 2
\node[stepbox=blue!6, below=0.3cm of s1] (s2)
  {\textbf{2.\ Sample join depth}\quad
   $N_{\text{join}} \sim U(1, K_{\text{join}}\!=\!3)$
   \quad$\Rightarrow$\quad $N_{\text{join}} = 3$};

% Step 3
\node[stepbox=orange!8, below=0.3cm of s2] (s3)
  {\textbf{3.\ Expand join edges} ($\times\,3$ iterations)\\[2pt]
   $\alpha_{\text{shape}} = 0.1$ (chain bias):
   each step favours the deepest frontier,\\
   suppressing root-attachment edges};

% Subgraph result
\node[grapharea, below=0.35cm of s3] (graph) {};
\node[font=\scriptsize\itshape, anchor=north west]
  at ([xshift=4pt, yshift=-4pt]graph.north west)
  {Resulting join subgraph (chain topology):};

% Draw the chain graph centered inside the box (lower half)
\node[rootnode] (t0) at ([xshift=-2.1cm, yshift=-0.2cm]graph.center) {\texttt{t}};
\node[tblnode, right=1.05cm of t0] (t1) {\texttt{ci}};
\node[tblnode, right=1.05cm of t1] (t2) {\texttt{n}};
\node[tblnode, right=1.05cm of t2] (t3) {\texttt{cn}};
\draw[arrow, blue!70, line width=0.9pt] (t0) --
  node[above, font=\tiny, yshift=1pt] {\textsf{movie\_id}} (t1);
\draw[arrow, blue!70, line width=0.9pt] (t1) --
  node[above, font=\tiny, yshift=1pt] {\textsf{person\_id}} (t2);
\draw[arrow, blue!70, line width=0.9pt] (t2) --
  node[above, font=\tiny, yshift=1pt] {\textsf{role\_id}} (t3);

% Arrows for Phase I
\draw[arrow] (p1hdr) -- (s1);
\draw[arrow] (s1) -- (s2);
\draw[arrow] (s2) -- (s3);
\draw[arrow] (s3) -- (graph);

% ======== CONNECTOR ========
\node[font=\normalsize\bfseries, text=gray!50,
      below=0.35cm of graph] (conn) {$\Downarrow$\;
      {\scriptsize\textnormal{blueprint
      $(\mathcal{T}_{\text{used}},\, J_q)$ passed to Phase~II}}};

% ======== PHASE II HEADER ========
\node[phaselabel=violet!15, below=0.35cm of conn] (p2hdr)
  {Phase~II: Semantic Injection + AST Assembly
   (Alg.~\ref{alg:phase2})};

% Step 4
\node[stepbox=violet!6, below=0.35cm of p2hdr] (s4)
  {\textbf{4.\ Column sampling + aggregation}\quad
   $P_{\text{agg}} = 1.0$\\[2pt]
   \texttt{t.title} $\to$ $C_{\text{select}}$,\;
   \texttt{cn.name} $\to$ $\mathrm{COUNT}(\cdot)$\quad
   $\Rightarrow$ \texttt{has\_agg} = True};

% Step 5
\node[stepbox=violet!6, below=0.3cm of s4] (s5)
  {\textbf{5.\ Predicate injection}\quad
   $P_{\text{where}} = 0.4$\quad (triggered)\\[2pt]
   Sampled: \texttt{t.production\_year > 2010}\quad
   from catalog domain};

% Step 6
\node[stepbox=violet!6, below=0.3cm of s5] (s6)
  {\textbf{6.\ AST compilation}\\[2pt]
   \texttt{has\_agg} $\wedge$ $C_{\text{group\_by}} \neq \emptyset$
   $\;\Rightarrow\;$ auto-append \texttt{GROUP BY t.title}\\
   Attach \texttt{ORDER BY}, \texttt{LIMIT} $\to$
   \texttt{CompileToSQLString}(AST)};

% SQL Output
\node[sqlbox, below=0.3cm of s6] (sql)
  {SELECT t.title, COUNT(cn.name)\\
   FROM title t\\
   \quad JOIN cast\_info ci ON t.id = ci.movie\_id\\
   \quad JOIN name n ON ci.person\_id = n.id\\
   \quad JOIN char\_name cn ON ci.person\_role\_id = cn.id\\
   WHERE t.production\_year > 2010\\
   GROUP BY t.title;};

% Output label
\node[outputlabel, below=0.3cm of sql] (out)
  {Output: $q_i$ appended to workload $\mathcal{Q}$};

% Arrows for Phase II
\draw[arrow] (p2hdr) -- (s4);
\draw[arrow] (s4) -- (s5);
\draw[arrow] (s5) -- (s6);
\draw[arrow] (s6) -- (sql);
\draw[arrow] (sql) -- (out);

\end{tikzpicture}%
}% end resizebox
\end{minipage}%
\hfill
\begin{minipage}[c]{0.36\textwidth}
\caption{\small Detailed walkthrough of a single SynQL iteration on the IMDb
schema.
\textbf{Phase~I} (steps 1--3): root table \texttt{title} is
  selected, join depth~3 is sampled, and three
  $\alpha_{\text{shape}}$-weighted edge expansions produce a chain
  subgraph with FK join conditions shown on each edge.
  \textbf{Phase~II} (steps 4--6): columns are sampled with full
  aggregation ($P_{\text{agg}}\!=\!1.0$), a year predicate is injected,
  and the AST compiler auto-appends \texttt{GROUP BY} before emitting
  the executable SQL query.}
\label{fig:walkthrough}
\end{minipage}
\end{figure}
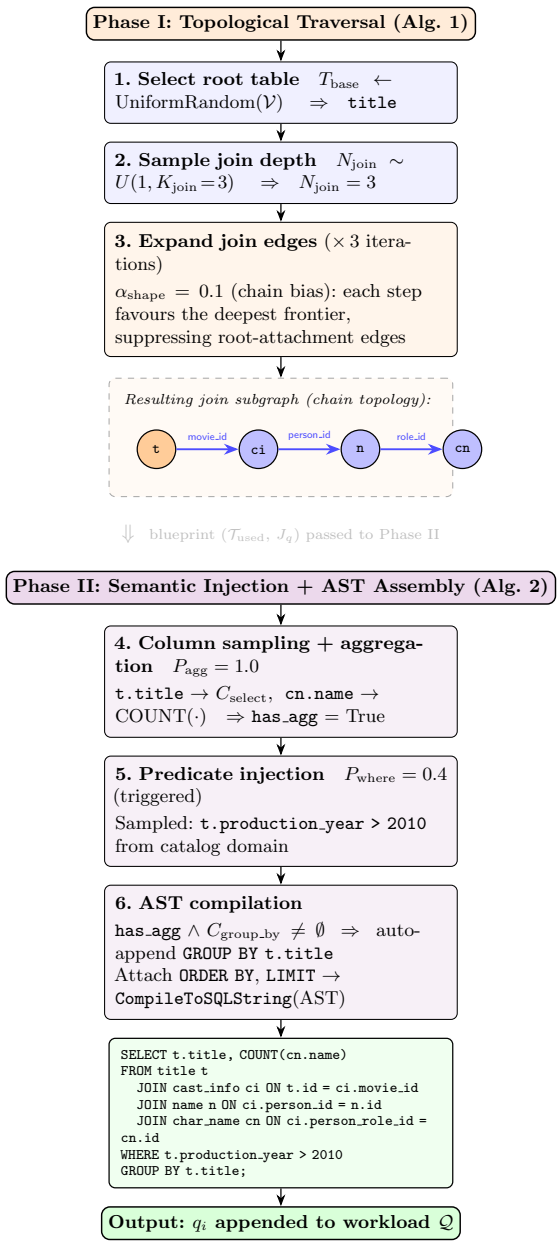

Figure~\ref{fig:walkthrough} traces a single iteration of this loop on
the IMDb schema, showing the concrete data flow from root-table
selection through join expansion, column sampling, and final AST
compilation into an executable SQL string.

% ---------- TikZ Detailed Walkthrough Diagram ----------
% Split across minipage: diagram left, caption right

% ------------------------------------------------------------------
\subsection{Configuration Vector $\Theta$: Optimizer Stress
            Dimensions}\label{subsec:theta}
% ------------------------------------------------------------------

Each parameter in~$\Theta$ targets a distinct, known failure mode of
query optimisers.  Table~\ref{tab:theta_summary} provides a compact
summary.

\paragraph{$\alpha_{\text{shape}}$ and $K_{\text{join}}$---Topology and
           Join Depth.}
$\alpha_{\text{shape}}$ controls the attachment bias of each new join
edge.  High values ($\alpha_{\text{shape}} \to 1$) produce Star schemas,
stressing fact-dimension join ordering---the regime where
Bao~\cite{bao2021} and Balsa~\cite{yang2022balsa} achieve their largest
gains over PostgreSQL.  Low values ($\alpha_{\text{shape}} \to 0$)
produce deep Chain queries, where estimation errors compound
multiplicatively---the scenario JOB-Complex~\cite{wehrstein2025job}
identifies as most damaging for current learned optimisers.
$K_{\text{join}}$ caps graph depth, bounding the join-ordering search
space per query.

\paragraph{$P_{\text{agg}}$---Analytical Intensity.}
High $P_{\text{agg}}$ generates OLAP-style queries with aggregation
functions, forcing the optimiser to choose between
\texttt{HashAggregate} and \texttt{GroupAggregate}.  This choice
interacts non-linearly with available memory (\texttt{work\_mem}) and
degree of parallelism---a dimension under-represented in standard fixed
templates.

\paragraph{$P_{\text{where}}$ and $K_{\text{pred}}$---Predicate
           Selectivity.}
These parameters control the density and width of \texttt{WHERE}
clauses, spanning the range from full-table scans to highly selective
point lookups.  Multi-predicate scenarios are precisely those where
histogram statistics are most prone to the independence-assumption
errors catalogued by Sun et al.~\cite{sun2021learnedcard}.

% ---------- Theta Summary Table ----------
\begin{table}[tb]
\centering
\caption{Configuration vector~$\Theta$: parameters and their effects
  on workload characteristics.}\label{tab:theta_summary}
\small
\begin{tabular}{@{} l l l p{4.6cm} @{}}
\toprule
\textbf{Parameter} & \textbf{Symbol} & \textbf{Range} &
  \textbf{Optimiser stress target} \\
\midrule
Topology bias    & $\alpha_{\text{shape}}$ & $[0,1]$
  & Star vs.\ Chain topology; join-order search space \\
Join depth limit & $K_{\text{join}}$       & $\mathbb{Z}^+$
  & Tables per query; cardinality error compounding \\
Aggregation prob.& $P_{\text{agg}}$        & $[0,1]$
  & OLAP intensity; Hash- vs.\ GroupAggregate choice \\
Predicate prob.  & $P_{\text{where}}$      & $[0,1]$
  & Selectivity range; independence-assumption errors \\
Predicate limit  & $K_{\text{pred}}$       & $\mathbb{Z}^+$
  & WHERE-clause width; multi-predicate estimation \\
\bottomrule
\end{tabular}
\end{table}

\section{Experiments}\label{sec:experiments}
%% ================================================================

All experiments were conducted on an Apple M1 workstation (8-core CPU, 16~GB
unified memory) running PostgreSQL~14~\cite{postgresql14} at default settings. Retaining the
default configuration is a deliberate choice: SynQL is entirely schema-driven
and generates corpora portable to any standard PostgreSQL deployment, aligning
with the environment-adaptation requirements noted in recent learned cost-model
studies~\cite{heinrich2025learned}. The SynQL framework algorithms and machine learning
pipeline were implemented in Python~3.9, using \texttt{psycopg2} for database
catalog interaction and \texttt{scikit-learn}~\cite{pedregosa2011sklearn}
for model training.

% ------------------------------------------------------------------
\subsection{Case Studies}\label{subsec:case_studies}
% ------------------------------------------------------------------

The $\Theta$ configuration vector exposes five independently tunable
dimensions.  The four case studies below isolate specific parameter
interactions to demonstrate the breadth of structural variation SynQL
can produce, covering all three topology types on both benchmark schemas.

\paragraph{Case Study~1: Star Join, Low Analytical Intensity (TPC-H).}
Setting $\alpha_{\text{shape}} = 0.9$, $K_{\text{join}} = 3$, and
$P_{\text{agg}} = 0.0$ directs SynQL to generate a wide, flat star
query with no aggregation---a projection-only scan of dimension tables
through the central \texttt{orders} fact table.  This configuration
stresses the join-ordering component of the optimiser: all three
dimension tables are equidistant from the root, so the planner must
evaluate six possible join orderings with no aggregation operator to
anchor its choice.

\begin{lstlisting}[
  language=SQL,
  caption={Star query on TPC-H ($\alpha_{\text{shape}}=0.9$,
           $K_{\text{join}}=3$, $P_{\text{agg}}=0.0$).},
  label={lst:cs1},
  basicstyle=\ttfamily\scriptsize,
  keywordstyle=\bfseries\color{blue!40!black},
  commentstyle=\itshape\color{green!40!black},
  numbers=left, numberstyle=\tiny\color{gray},
  stepnumber=1, numbersep=8pt,
  frame=lines, breaklines=true,
  showstringspaces=false, tabsize=2
]
-- Case Study 1: TPC-H star, no aggregation
SELECT o.orderdate, c.mktsegment, s.name, n.name
FROM orders o
  JOIN customer c ON o.custkey = c.custkey
  JOIN supplier s ON o.orderkey = s.suppkey
  JOIN nation   n ON c.nationkey = n.nationkey
WHERE o.orderdate > '1995-01-01';
\end{lstlisting}

\paragraph{Case Study~2: Chain Join, High Analytical Intensity (TPC-H).}
Setting $\alpha_{\text{shape}} = 0.05$, $K_{\text{join}} = 4$, and
$P_{\text{agg}} = 0.8$ forces a deep five-table chain with heavy OLAP
aggregation.  The sequential join path means cardinality estimation
errors compound multiplicatively at each step, and the dense
\texttt{SUM}/\texttt{AVG} aggregations require the planner to make
memory-sensitive decisions between \texttt{HashAggregate} and
\texttt{GroupAggregate}---exactly the interaction under-represented in
TPC-H's 22 fixed templates.

\begin{lstlisting}[
  language=SQL,
  caption={Chain query on TPC-H ($\alpha_{\text{shape}}=0.05$,
           $K_{\text{join}}=4$, $P_{\text{agg}}=0.8$).},
  label={lst:cs2},
  basicstyle=\ttfamily\scriptsize,
  keywordstyle=\bfseries\color{blue!40!black},
  commentstyle=\itshape\color{green!40!black},
  numbers=left, numberstyle=\tiny\color{gray},
  stepnumber=1, numbersep=8pt,
  frame=lines, breaklines=true,
  showstringspaces=false, tabsize=2
]
-- Case Study 2: TPC-H chain, high aggregation
SELECT n.name, r.name,
       SUM(l.extendedprice) AS total_revenue,
       AVG(l.discount)      AS avg_discount
FROM lineitem l
  JOIN orders   o ON l.orderkey   = o.orderkey
  JOIN customer c ON o.custkey    = c.custkey
  JOIN nation   n ON c.nationkey  = n.nationkey
  JOIN region   r ON n.regionkey  = r.regionkey
WHERE l.shipdate > '1994-01-01'
GROUP BY n.name, r.name
ORDER BY total_revenue DESC;
\end{lstlisting}

\paragraph{Case Study~3: Fork Join, Multi-Predicate Selectivity (IMDb).}
Setting $\alpha_{\text{shape}} = 0.5$, $K_{\text{join}} = 3$,
$P_{\text{agg}} = 0.3$, $P_{\text{where}} = 1.0$, and $K_{\text{pred}}
= 3$ produces a branching fork topology on IMDb with multiple simultaneous
filter predicates.  The fork structure requires the planner to manage two
independent join sub-trees hanging from the root, while three concurrent
predicates expose the histogram independence-assumption errors documented
by Sun et al.~\cite{sun2021learnedcard}.  IMDb's extreme data skew on
\texttt{title.production\_year} amplifies this effect, making this
configuration particularly informative for training cardinality estimators.

\begin{lstlisting}[
  language=SQL,
  caption={Fork query on IMDb ($\alpha_{\text{shape}}=0.5$,
           $K_{\text{join}}=3$, $P_{\text{where}}=1.0$,
           $K_{\text{pred}}=3$).},
  label={lst:cs3},
  basicstyle=\ttfamily\scriptsize,
  keywordstyle=\bfseries\color{blue!40!black},
  numbers=left, numberstyle=\tiny\color{gray},
  stepnumber=1, numbersep=8pt,
  frame=lines, breaklines=true,
  showstringspaces=false, tabsize=2
]
-- Case Study 3: IMDb fork, multi-predicate
SELECT t.title, mk.keyword, AVG(mi.info) AS avg_info
FROM title t
  JOIN movie_keyword mk ON t.id = mk.movie_id
  JOIN movie_info    mi ON t.id = mi.movie_id
  JOIN keyword        k ON mk.keyword_id = k.id
WHERE t.production_year > 2005
  AND t.kind_id = 1
  AND mk.keyword_id < 5000
GROUP BY t.title, mk.keyword;
\end{lstlisting}

\paragraph{Case Study~4: Deep Chain, Fully Analytical (IMDb).}
Setting $\alpha_{\text{shape}} = 0.1$, $K_{\text{join}} = 4$,
$P_{\text{agg}} = 1.0$, and $P_{\text{where}} = 0.0$ generates a
five-table linear chain on IMDb with every numeric column aggregated and no
\texttt{WHERE} clause.  The absence of predicates means the planner receives
no selectivity signals to guide join reordering---all intermediate result
sizes must be estimated from base-table statistics alone.  This worst-case
estimation scenario is compounded by IMDb's skewed row distributions across
the \texttt{cast\_info}--\texttt{name}--\texttt{char\_name} spine, and the
full aggregation forces a \texttt{GroupAggregate} over the entire cross
product before any filtering can reduce the working set.

\begin{lstlisting}[
  language=SQL,
  caption={IMDb deep Chain query ($\alpha_{\text{shape}}=0.1$,
           $K_{\text{join}}=4$, $P_{\text{agg}}=1.0$,
           $P_{\text{where}}=0.0$).},
  label={lst:cs4},
  basicstyle=\ttfamily\scriptsize,
  keywordstyle=\bfseries\color{blue!40!black},
  commentstyle=\itshape\color{green!40!black},
  numbers=left, numberstyle=\tiny\color{gray},
  stepnumber=1, numbersep=8pt,
  frame=lines, breaklines=true,
  showstringspaces=false, tabsize=2
]
-- CS4: IMDb deep chain, fully analytical, no predicates
SELECT t.title,
       SUM(mi.info)   AS total_info,
       AVG(mi.info)   AS avg_info,
       COUNT(cn.name) AS role_count
FROM title t
  JOIN cast_info ci ON t.id              = ci.movie_id
  JOIN name      n  ON ci.person_id      = n.id
  JOIN char_name cn ON ci.person_role_id = cn.id
  JOIN movie_info mi ON t.id             = mi.movie_id
GROUP BY t.title
ORDER BY role_count DESC
LIMIT 100;
\end{lstlisting}

Table~\ref{tab:cs_summary} summarises all four case studies, highlighting how
each targets a distinct optimiser stress dimension absent or under-represented
in standard benchmark template sets.

\begin{table}[tb]
\centering
\caption{Summary of case studies and their optimiser stress targets.}
\label{tab:cs_summary}
\small
\begin{tabular}{@{} c l c c c p{3.2cm} @{}}
\toprule
\textbf{CS} & \textbf{Schema} & $\alpha_{\text{shape}}$
  & $K_{\text{join}}$ & $P_{\text{agg}}$ & \textbf{Stress target} \\
\midrule
1 & TPC-H & 0.90 & 3 & 0.0
  & Join ordering without aggregate anchor \\
2 & TPC-H & 0.05 & 4 & 0.8
  & Compounding chain errors + OLAP agg.\ \\
3 & IMDb  & 0.50 & 3 & 0.3
  & Multi-predicate independence assumption \\
4 & IMDb  & 0.10 & 4 & 1.0
  & Full aggregation, no selectivity signal \\
\botrule
\end{tabular}
\end{table}

% ------------------------------------------------------------------
\subsection{Workload Generation Using Benchmark
            Templates}\label{subsec:workload_gen}
% ------------------------------------------------------------------

\subsubsection{Target Schemas.}\label{subsubsec:schemas}

We validated SynQL against two schemas with strongly contrasting structural
properties.

\textbf{TPC-H}~\cite{poess2000tpc} is an 8-table normalised retail data
warehouse schema representative of OLAP workloads.  Its core structure
centres on a \texttt{lineitem}--\texttt{orders}--\texttt{customer} spine,
augmented by three dimension tables (\texttt{supplier}, \texttt{part},
\texttt{partsupp}) and a two-level geographical hierarchy
(\texttt{nation}, \texttt{region}).  The schema is acyclic and
relatively shallow (maximum FK path depth of~4), making it a controlled
baseline for assessing topological diversity: any workload generator that
defaults to hub-and-spoke star patterns will over-weight joins through
the \texttt{orders} fact table while neglecting the supplier and parts
sub-graphs.  SynQL's $\alpha_{\text{shape}}$ parameter explicitly
breaks this bias by assigning traversal weights that promote chain and
fork exploration across the full FK graph.

\textbf{IMDb (Join Order Benchmark)}~\cite{leis2015good} is a 21-table
real-world schema derived from the Internet Movie Database, characterised
by extreme data skew, dense many-to-many relationships (e.g.,
\texttt{title}--\texttt{cast\_info}--\texttt{name}), and cyclical FK
graphs that prevent naive tree-traversal strategies from covering the
schema uniformly.  The JOB benchmark~\cite{leis2015good,leis2024still}
uses 113 hand-crafted queries over this schema to expose cardinality
estimation failures in modern optimisers; our SynQL-generated corpus
supplements these templates with 10,000 structurally diverse queries
that systematically vary join depth, topology, and predicate selectivity.
IMDb's hub tables (\texttt{title}, \texttt{cast\_info}) attract
high-degree FK connections, meaning that a star-biased generator will
over-sample these nodes.  SynQL counters this by tracking
$\mathcal{T}_{\text{used}}$ and restricting re-entry, ensuring
schema-wide table coverage even in the presence of dominant hub nodes.

\subsubsection{Workload Generation Configuration.}\label{subsubsec:config}

SynQL synthesised 20,000 queries (10,000 per schema) using the balanced
configuration shown in Table~\ref{tab:config}.  The central value
$\alpha_{\text{shape}} = 0.5$ was chosen to avoid skewing the generated
distribution toward any single topology, producing a roughly balanced
mix of Star, Chain, and Fork queries whose precise proportions are
reported in Section~\ref{subsec:diversity}.  $K_{\text{join}} = 3$
bounds query width to at most four tables, matching the depth range of
TPC-H's most complex standard templates while remaining tractable for
PostgreSQL's join-order planner.  The predicate parameters
($P_{\text{where}} = 0.4$, $K_{\text{pred}} = 3$) were selected to
mirror the selectivity profile of the JOB workload, where approximately
40\% of queries carry multi-column \texttt{WHERE} clauses.

\begin{table}[tb]
\centering
\caption{SynQL configuration $\Theta$ for benchmark workload generation.}
\label{tab:config}
\small
\begin{tabular}{@{}llcl@{}}
\toprule
\textbf{Parameter} & \textbf{Symbol} & \textbf{Value}
  & \textbf{Effect} \\
\midrule
Join depth limit  & $K_{\text{join}}$       & 3   & Up to 4 tables per query \\
Aggregation prob. & $P_{\text{agg}}$        & 0.2 & 20\% OLAP-style queries  \\
Predicate prob.   & $P_{\text{where}}$      & 0.4 & 40\% filtered queries    \\
Predicate limit   & $K_{\text{pred}}$       & 3   & Max \texttt{WHERE} conditions \\
Topology bias     & $\alpha_{\text{shape}}$ & 0.5 & Balanced Star/Chain/Fork \\
\botrule
\end{tabular}
\end{table}

% ------------------------------------------------------------------
\subsection{Query Execution Time Prediction}\label{subsec:qet_prediction}
% ------------------------------------------------------------------

\subsubsection{Training Phase: Feature Engineering and Model Training.}\label{subsubsec:features}

Ground-truth execution-time labels were collected by running the synthetic
workload generated as described in section \ref{subsec:workload_gen} through PostgreSQL's \texttt{EXPLAIN ANALYZE} command. Input features
were derived exclusively from pre-execution planner estimates, so that no
post-execution statistics leak into the feature vector~\cite{sun2019end},
preserving each model's utility as a genuine pre-execution predictor.
Table~\ref{tab:features} enumerates the complete feature set.

\begin{table}[tb]
\centering
\caption{Feature vector for QET prediction (21 features). All features
  are extracted from PostgreSQL's \texttt{EXPLAIN} output
  (pre-execution).}
\label{tab:features}
\small
\begin{tabular}{@{} c l p{4.8cm} @{}}
\toprule
\textbf{\#} & \textbf{Feature} & \textbf{Description} \\
\midrule
\multicolumn{3}{@{}l}{\textit{Planner cost estimates}} \\
1  & \texttt{plan\_total\_cost}   & Estimated total cost of the root plan node \\
2  & \texttt{plan\_startup\_cost} & Estimated cost to return the first row \\
3  & \texttt{plan\_rows}          & Estimated row count at the root node \\
4  & \texttt{plan\_width}         & Estimated average row width (bytes) \\
5  & \texttt{max\_plan\_rows}     & Maximum \texttt{plan\_rows} across all plan nodes \\
\midrule
\multicolumn{3}{@{}l}{\textit{Plan structural features}} \\
6  & \texttt{num\_plan\_nodes}    & Total number of nodes in the plan tree \\
7  & \texttt{plan\_depth}         & Depth (height) of the plan tree \\
8  & \texttt{num\_joins}          & Number of join operators \\
9  & \texttt{num\_relations}      & Number of base relations accessed \\
10 & \texttt{num\_predicates}     & Number of filter / join predicates \\
\midrule
\multicolumn{3}{@{}l}{\textit{Operator-type counts}} \\
11 & \texttt{num\_seq\_scan}      & Sequential Scan nodes \\
12 & \texttt{num\_index\_scan}    & Index Scan / Index Only Scan nodes \\
13 & \texttt{num\_bitmap\_scan}   & Bitmap Heap / Index Scan nodes \\
14 & \texttt{num\_hash\_join}     & Hash Join nodes \\
15 & \texttt{num\_merge\_join}    & Merge Join nodes \\
16 & \texttt{num\_nested\_loop}   & Nested Loop join nodes \\
17 & \texttt{num\_aggregate}      & Aggregate / GroupAggregate nodes \\
18 & \texttt{num\_sort}           & Sort nodes \\
19 & \texttt{num\_limit}          & Limit nodes \\
20 & \texttt{num\_materialize}    & Materialize nodes \\
21 & \texttt{num\_gather}         & Gather / Gather Merge (parallel) nodes \\
\botrule
\end{tabular}
\end{table}

The dataset was split 80/20 for
training and testing. We evaluated three tree-based
ensembles---Random Forest~\cite{breiman2001random},
XGBoost~\cite{chen2016xgboost}, and Gradient
Boosting~\cite{friedman2001greedy}---selected for their
ability to model non-linear operator interactions, as validated by the
industrial Stage predictor~\cite{wu2024stage}. This choice of tree-based
ensembles for tabular performance prediction aligns with recent findings that
gradient-boosted models consistently outperform deep architectures on
structured feature sets~\cite{zhang2024performance}. Performance is reported
via RMSE, MAE, and $R^{2}$.

\subsubsection{Prediction Phase:}\label{subsubsec:prediction}

Having established that SynQL produces structurally diverse, schema-valid
workloads (Section~\ref{subsec:workload_gen}), we now evaluate whether
those workloads constitute effective training data for learned QET
predictors.  The experimental protocol follows the standard
pre-execution prediction paradigm~\cite{sun2019end}: features are
extracted from PostgreSQL's \texttt{EXPLAIN} output (no post-execution
statistics), and the target is the median wall-clock execution time
(P50) measured over five repeated runs to suppress OS-level jitter.
Execution time targets are $\log_{1p}$-transformed before training to
compress the heavy-tailed runtime distribution, with the inverse
transform applied at inference.

The complete results are reported in Section~\ref{sec:results}.  Here
we note three design decisions that distinguish this evaluation from
prior synthetic-workload studies.  First, the strict pre-execution
feature constraint ensures that reported $R^{2}$ values reflect genuine
\emph{predictive} performance, not post-hoc curve fitting.  Second, the
stratified 5-fold cross-validation protocol guards against topology
imbalance inflating aggregate metrics.  Third, the cross-topology
transfer experiment (Section~\ref{subsec:per-topo}) directly tests
whether topological diversity in the training corpus---SynQL's core
contribution---translates into robustness across query shapes.

%% ================================================================
\clearpage
\section{Results and Discussion}\label{sec:results}
%% ================================================================

\subsection{Workload Characterisation and Topological
            Diversity}\label{subsec:diversity}

A primary design goal of SynQL is to prevent the topological collapse observed
in LLM-based generators. We quantify structural diversity using Shannon Entropy
($H$) over the topological distribution (Star, Chain, Fork).

As shown in Table~\ref{tab:shape-distribution}, the SynQL-generated TPC-H
workload achieves $H = 1.53$~bits, closely approaching the theoretical maximum
for three categories ($\log_2 3 \approx 1.58$~bits). SynQL produced a
well-balanced mix of Chain (43.8\%), Star (32.9\%), and Fork (23.3\%)
topologies. For the IMDb schema, the generator appropriately reflected the
database's inherent hub-and-spoke connectivity, yielding a Star-dominant
distribution (53.5\%) and $H = 1.34$~bits. Both values substantially exceed
the structural diversity afforded by the 22 static templates of the standard
TPC-H benchmark.

\begin{table}[tb]
\centering
\caption{Topological distribution of the 20,000-query SynQL corpus.}
\label{tab:shape-distribution}
\small
\begin{tabular}{lcc}
\toprule
\textbf{Topology} & \textbf{TPC-H (\%)} & \textbf{IMDb (\%)} \\
\midrule
Chain             & 43.82 & 12.82 \\
Star              & 32.89 & 53.50 \\
Fork / Two-Table  & 23.29 & 33.68 \\
\midrule
\textbf{Entropy $H$} & \textbf{1.53 bits} & \textbf{1.34 bits} \\
\botrule
\end{tabular}
\end{table}

\subsection{Predictive Accuracy and Deployment
            Viability}\label{subsec:accuracy}

Table~\ref{tab:results} reports test-set performance of the learned QET
predictors. To ensure statistical robustness, we performed stratified 5-fold
cross-validation; the table shows results from the 80/20 held-out split,
which fell within one standard deviation of the cross-validated means in all
cases.

On TPC-H, all three ensembles achieved $R^{2} > 0.98$, with XGBoost
reaching $R^{2} = 0.987$ ($\text{CV}: 0.980 \pm 0.004$). On IMDb---a
substantially harder schema due to extreme data skew and 27\% query
timeouts---Random Forest achieved the highest $R^{2} = 0.824$
($\text{CV}: 0.791 \pm 0.035$). The performance gap between schemas
is expected: IMDb's dense many-to-many relationships and highly variable
execution times create a more challenging prediction landscape, consistent
with the difficulties reported by the JOB
benchmark~\cite{leis2015good,leis2024still}. All three models executed
with sub-millisecond inference latency, satisfying the strict overhead
constraints required for deployment on a live query optimiser's critical
path.

\begin{table}[tb]
\centering
\caption{Model performance on held-out test sets (80/20 split).
  5-fold CV $R^{2}$ shown as mean $\pm$ std.}
\label{tab:results}
\small
\begin{tabular}{@{}lcccccc@{}}
\toprule
& \multicolumn{3}{c}{\textbf{TPC-H}}
& \multicolumn{3}{c}{\textbf{IMDb}} \\
\cmidrule(lr){2-4}\cmidrule(lr){5-7}
\textbf{Model}
  & \textbf{RMSE} & \textbf{MAE} & $\mathbf{R^{2}}$
  & \textbf{RMSE} & \textbf{MAE} & $\mathbf{R^{2}}$ \\
\midrule
Random Forest  & 0.27 & 0.12 & 0.986 & 0.80 & 0.43 & 0.824 \\
XGBoost        & 0.27 & 0.13 & 0.987 & 0.86 & 0.46 & 0.794 \\
Gradient Boost & 0.27 & 0.12 & 0.987 & 0.80 & 0.42 & 0.822 \\
\botrule
\end{tabular}
\end{table}

\subsection{Per-Topology Prediction Analysis}\label{subsec:per-topo}

To validate that SynQL's topological diversity translates into effective
training signal across \emph{all} join shapes, we classify each generated
query's topology by parsing its join graph from the SQL text: a query is
labelled \textbf{Star} if all joins attach to the root table (including
two-table joins as degenerate stars), \textbf{Chain} if each successive
table joins only to the previous one, and \textbf{Fork} otherwise.
Table~\ref{tab:per-topo} disaggregates XGBoost's predictions by topology
class on the held-out test set.

\begin{table}[tb]
\centering
\caption{Per-topology XGBoost performance on held-out test sets.
  Topology labels are derived from the SQL join graph.}
\label{tab:per-topo}
\small
\begin{tabular}{@{}lcccccc@{}}
\toprule
& \multicolumn{3}{c}{\textbf{TPC-H}}
& \multicolumn{3}{c}{\textbf{IMDb}} \\
\cmidrule(lr){2-4}\cmidrule(lr){5-7}
\textbf{Topology}
  & \textbf{RMSE} & \textbf{MAE} & $\mathbf{R^{2}}$
  & \textbf{RMSE} & \textbf{MAE} & $\mathbf{R^{2}}$ \\
\midrule
Star   & 0.24 & 0.12 & 0.990
       & 1.08 & 0.59 & 0.671 \\
Chain  & 0.30 & 0.14 & 0.982
       & 0.77 & 0.28 & 0.862 \\
Fork   & 0.24 & 0.12 & 0.976
       & 0.80 & 0.44 & 0.818 \\
\midrule
All    & 0.27 & 0.13 & 0.987
       & 0.86 & 0.46 & 0.794 \\
\botrule
\end{tabular}
\end{table}

Each benchmark uses a held-out test set of 2\,000 queries. On TPC-H,
performance is uniformly high across all topologies ($R^{2} \geq
0.976$), with Star queries being easiest ($R^{2} = 0.990$) and Chain queries
hardest ($R^{2} = 0.982$)---consistent with chains producing compounding
estimation errors.  On IMDb, Chain queries achieve the highest
$R^{2} = 0.862$ despite being the rarest topology (only 68 test queries),
while Star queries are harder ($R^{2} = 0.671$) due to IMDb's hub tables
producing highly variable cardinalities.  Fork queries dominate the IMDb
test set (1\,560 of 2\,000) yet still reach $R^{2} = 0.818$, indicating
that the model generalises well even for the most frequent topology.

The results are striking.  On TPC-H, a model trained exclusively on
Fork queries yields $R^{2} = -0.142$ when tested on Star
queries---\emph{worse than predicting the mean}.  On IMDb, the effect
is even more pronounced: Star-trained models produce
$R^{2} = -1.313$ on Chain queries.  In every case, the mixed-topology
model (``All'', bolded) matches or exceeds the best single-topology
model on every test partition.  This directly validates SynQL's core
design principle: topologically diverse training corpora are not merely
desirable but \emph{necessary} for robust cost prediction across the
full range of query structures.

\subsection{SynQL vs.\ LLM-Based Workload
            Generation}\label{subsec:comparison}

SynQL and LLM-based SQL generators address fundamentally different tasks---
SynQL performs \emph{schema-driven workload synthesis} for training data
generation, whereas systems evaluated by Spider~2.0~\cite{lei2024spider2} and
BIRD~\cite{li2023bird} perform \emph{natural-language-to-SQL translation}.
A direct numerical comparison of success rates is therefore not meaningful.
Nevertheless, the failure modes identified in these benchmarks are directly
relevant to workload synthesis, because any generator that produces invalid or
structurally homogeneous queries is unsuitable for training learned optimisers.

Three properties distinguish SynQL's constructive approach from LLM-based
generation.

\paragraph{Schema Validity.}
Spider~2.0 reports that GPT-4o's success rate drops sharply on enterprise
schemas due to schema hallucination and dialect confusion, while BIRD
observes execution accuracy below 40\% on complex schemas. By contrast,
SynQL deterministically bounds generation to the database's live foreign-key
graph and enforces strict AST compliance
(Section~\ref{sec:framework}), achieving 100\% schema validity and zero
syntax errors across the entire 20,000-query corpus.

\paragraph{Topological Diversity.}
LLMs probabilistically collapse toward simple star
joins~\cite{zhou2025parrotbenchmarkevaluatingllms} because hub-and-spoke
patterns dominate their pre-training corpora. The
$\alpha_{\text{shape}}$ parameter provides mathematical control over topology,
enabling SynQL to synthesise deep join chains (43.8\% on TPC-H) that expose
the compound estimation errors causing performance regressions in learned
optimisers~\cite{wehrstein2025job}.

\paragraph{Deterministic Reproducibility.}
Given identical $\Theta$ and random seed, SynQL produces byte-identical
workloads, enabling controlled ablation studies. LLM outputs are inherently
stochastic and sensitive to prompt phrasing, making experimental
reproducibility difficult.

%% ================================================================
\section{Limitations and Threats to Validity}\label{sec:limitations}
%% ================================================================

We identify the following limitations of the current work, which also
define concrete directions for improvement.

\paragraph{SQL Coverage.}
SynQL is designed as a \emph{foundational framework} targeting the core SQL
fragment that dominates analytical workloads: multi-table \texttt{SELECT}
queries with inner joins, projections, optional aggregations, and range
predicates. This fragment accounts for the vast majority of queries in
standard benchmarks (TPC-H, JOB) and captures the join-topology and
selectivity dimensions most critical for stressing learned
optimisers~\cite{wehrstein2025job}. SynQL does not yet support correlated
subqueries, \texttt{EXISTS}/\texttt{IN} clauses, Common Table Expressions
(CTEs), set operations (\texttt{UNION}, \texttt{INTERSECT}), or
\texttt{HAVING} clauses. Crucially, however, SynQL's AST-based architecture
is \emph{extensible by design}: adding new clause types requires
implementing additional \texttt{AttachNode} rules in Phase~II without
modifying the topological traversal of Phase~I. We view the current work as
establishing the core generation paradigm, with richer SQL constructs as a
natural and modular extension.

\paragraph{Database Engine Scope and Feature Portability.}
All experiments were conducted on PostgreSQL~14~\cite{postgresql14}.
It is important to distinguish two layers of engine dependence in SynQL's
design.

The \emph{query generation} layer (Phase~I and Phase~II) is fully
engine-agnostic: it operates on the relational schema graph and emits
standard SQL\@.  The generated queries are syntactically portable to any
SQL-compliant engine---we verified that the TPC-H corpus parses without
errors on both PostgreSQL and SQLite.

The \emph{feature extraction} layer (Table~\ref{tab:features}), however,
is tied to PostgreSQL's \texttt{EXPLAIN} output format.  Adapting SynQL's
ML pipeline to other engines requires mapping the 21 plan-level features
to engine-specific equivalents.  This is feasible for most targets:
Spark SQL exposes analogous plan metadata via
\texttt{EXPLAIN EXTENDED}, Snowflake provides query profiles with
operator-level statistics, and MySQL's \texttt{EXPLAIN ANALYZE}
(available since 8.0) reports comparable operator counts and cost
estimates.  The core feature categories---planner cost estimates, plan
structural features, and operator-type counts---have natural
counterparts in all major engines, though the specific operator
vocabulary differs (e.g., Snowflake uses \texttt{TableScan} rather than
\texttt{SeqScan}).  Validating this cross-engine feature mapping and
measuring whether SynQL-trained cost models transfer across engines
remains an important direction for future work.

\paragraph{Schema Diversity.}
The evaluation covers two schemas (TPC-H and IMDb). While these represent
contrasting structural properties (normalised warehouse vs.\ skewed
real-world graph), the results may not generalise to schemas with
substantially different characteristics, such as deeply nested
hierarchies or very large table counts ($>$100 tables).

\paragraph{Evaluation on Synthetic Data and Production Transfer.}
The $R^{2}$ values reported in Section~\ref{subsec:accuracy} reflect
prediction accuracy on held-out \emph{synthetic} test queries generated
by SynQL itself. This evaluation design is deliberate: it establishes
that the synthetic corpus provides a training signal of sufficient quality
and diversity for accurate cost modelling---a necessary prerequisite
before any production deployment. However, it does not directly measure
\emph{transfer} to real production workloads, which may exhibit
distributional characteristics absent from the current generator (e.g.,
deeply nested subqueries, user-defined functions, highly skewed parameter
distributions).

We note that the production-transfer gap is a challenge shared by all
synthetic workload generators, including fixed benchmarks like TPC-H.
The key advantage of SynQL in this context is its parametric
controllability: practitioners can tune $\Theta$ to approximate known
characteristics of their production workload (e.g., high chain depth for
OLTP-heavy systems, high $P_{\text{agg}}$ for data-warehouse queries)
without exposing proprietary SQL text. Validating this transfer pathway
on anonymised production traces (e.g., Redset execution statistics) is a
high-priority direction for future work.

\paragraph{Topological Entropy Granularity.}
Shannon entropy is computed over three coarse topology categories (Star,
Chain, Fork). This metric does not capture within-category variation
(e.g., chain length distribution). A finer-grained diversity metric,
such as graph-edit-distance-based measures, would provide a more
nuanced characterisation.

%% ================================================================
\section{Conclusion}\label{sec:conclusion}
%% ================================================================

We introduced SynQL, a deterministic workload synthesis framework designed to
overcome the training-data scarcity that bottlenecks the deployment of learned
database systems. By replacing probabilistic text generation with a two-phase
constructive pipeline---schema-driven topological graph traversal followed by
strict AST assembly---SynQL mechanically eliminates the schema hallucinations
and syntax errors that severely limit LLM-based generators.

Our evaluation demonstrates that the topological bias parameter
$\alpha_{\text{shape}}$ effectively prevents mode collapse: SynQL generated a
20,000-query corpus across TPC-H and IMDb achieving near-maximal topological
entropy ($H = 1.53$~bits), natively producing the deep join chains and complex
fork topologies absent from standard benchmarks. Tree-based cost models trained
on this synthetic corpus achieved accurate execution time predictions
($R^{2} \geq 0.79$ on held-out synthetic test sets, reaching $0.99$ on
TPC-H) with sub-millisecond inference latency. Crucially, our cross-topology
transfer experiment (Section~\ref{subsec:per-topo}) demonstrates that models
trained on a single topology \emph{fail catastrophically} on other topologies
(negative $R^{2}$), while the mixed-topology corpus consistently yields the
best performance across all shapes---directly validating SynQL's core design
principle that topological diversity in training data is not merely
desirable but necessary.

Four primary directions remain for future work. First, expanding the AST
compiler to support correlated subqueries, \texttt{EXISTS} clauses, and Common
Table Expressions (CTEs) would directly address the optimizer failure modes
highlighted by JOB-Complex~\cite{wehrstein2025job}; SynQL's modular
architecture makes this a natural extension without modifying Phase~I.
Second, validating the \emph{synthetic-to-production transfer} pathway---by
training models on SynQL-generated corpora and evaluating on anonymised
production traces (e.g., Redset execution statistics)---would establish the
strategic value of controllable synthesis for industrial deployment.
Third, the structured relational trees produced during SynQL's assembly phase
are naturally suited for pre-training Graph Neural Networks for plan-cost
estimation~\cite{sun2019end}, offering richer representations than flat feature
vectors. Finally, abstracting the AST layer to support multi-dialect generation
(Spark SQL, Snowflake, BigQuery) and validating cross-engine feature mapping
would cement SynQL's industrial value: if a cost model trained on
PostgreSQL-executed SynQL queries transfers to Spark or Snowflake plan
features with minimal accuracy loss, it would demonstrate that synthetic
workloads can serve as a universal training substrate across heterogeneous
enterprise environments---the scenario documented in the Spider~2.0
challenge~\cite{lei2024spider2} and increasingly demanded by cloud-native
data platforms~\cite{heinrich2025learned}.

%% ================================================================
%% BIBLIOGRAPHY
%% ================================================================
\bibliographystyle{unsrtnat}
\bibliography{papers}

\end{document}